\documentclass[12pt]{iopart}

\newcommand{\x} {{\bf x}}
\newcommand{\xk} {{\bf x}^{(k)}}
\newcommand{\xl} {{\bf x}^{(l)}}
\newcommand{\xki} {x^{(k)}_i}

\newcommand{\w} {{\bf w}}
\newcommand{\wx} {{\bf w} \cdot {\bf x}}
\newcommand{\wxk} {{\bf w} \cdot \xk }

\usepackage{iopams, color}  
\usepackage{color}  

\begin{document}
\bibliographystyle{unsrt}

\title{Regularizing Portfolio Optimization.}

\author{Susanne Still}
\address{Information and Computer Sciences, University of Hawaii at M$\bar{\rm a}$noa, Honolulu, Hawaii, USA} 
\ead{sstill@hawaii.edu} 
\author{Imre Kondor}
\address{Collegium Budapest--Institute for Advanced Study and Department of Physics of Complex Systems, E{\"o}tv{\"o}s University, Budapest, Hungary}
\ead{kondor@colbud.hu}
\begin{abstract}

\noindent  The optimization of large portfolios displays an inherent instability to estimation error. This poses a fundamental problem, because solutions that are not stable under sample fluctuations may look optimal for a given sample, but are, in effect, very far from optimal with respect to the average risk. In this paper, we approach the problem from the point of view of statistical learning theory. The occurrence of the instability is intimately related to over-fitting which can be avoided using known regularization methods. We show how regularized portfolio optimization with the expected shortfall as a risk measure is related to support vector regression. The budget constraint dictates a modification. We present the resulting optimization problem and discuss the solution.  The L2 norm of the weight vector is used as a regularizer,  which corresponds to a diversification ``pressure". This means that diversification, besides counteracting downward fluctuations in some assets by upward fluctuations in others, is also crucial because it improves the stability of the solution. The approach we provide here allows for the simultaneous treatment of optimization and diversification in one framework that enables the investor to trade-off between the two, depending on the size of the available data set.
\end{abstract}
\maketitle

\section{Introduction}
Markowitz' portfolio selection theory \cite{Markowitz52,Markowitz59} is one of the pillars of theoretical finance. It has greatly influenced the thinking and practice in investment, capital allocation, index tracking, and a number of other fields. Its two major ingredients are (i) seeking a trade-off between risk and reward, and (ii) exploiting the cancellation between fluctuations of (anti-)correlated assets. In the original formulation of the theory, the underlying process was assumed to be multivariate normal. Accordingly, reward was measured in terms of the expected return, risk in terms of the variance of the portfolio.

The fundamental problem of this scheme (shared by all the other variants that have been introduced since) is that the characteristics of the underlying process generating the distribution of asset prices are not known in practice, and therefore averages are replaced by sums over the available sample. This procedure is well justified as long as the sample size, $T$ (i.e. the length of the available time series for each item), is sufficiently large compared to the size of the portfolio, $N$ (i.e. the number of items). In that limit, sample averages asymptotically converge to the true average due to the central limit theorem.

Unfortunately, the nature of portfolio selection is not compatible with this limit. Institutional portfolios are large, with $N$'s in the range of hundreds or thousands, while considerations of transaction costs and non-stationarity limit the number of available data points to a couple of hundreds at most. Therefore, portfolio selection works in a region, where $N$ and $T$ are, at best, of the same order of magnitude. This, however, is not the realm of classical statistical methods. Portfolio optimization is rather closer to a situation which, by borrowing a term from statistical physics, might be termed the ``thermodynamic limit", where $N$ and $T$ tend to infinity such that their ratio remains fixed.

It is evident that portfolio theory struggles with the same fundamental difficulty that is underlying basically every complex modeling and optimization task: the high number of dimensions and the insufficient amount of information available about the system. This difficulty has been around in portfolio selection from the early days and a plethora of methods have been proposed to cope with it, e.g. single and multi-factor models \cite{eltongruber}, Bayesian estimators \cite{jobson1979,jorion1986,frost_savarino,safePO,Jagannathan2003,LedoitWolf2003, LedoitWolf2004,LedoitWolfHoney,DeMiguel2007,Garlappi2007,Golosnoy2007,Kan2007,frahm_memmel,DeMiguel2009}, or, more recently, tools borrowed from random matrix theory \cite{Laloux1999,Plerou1999,Laloux2000,Plerou2002,burda,Potters2005}. In the thermodynamic regime, estimation errors are large, sample to sample fluctuations are huge, results obtained from one sample do not generalize well and can be quite misleading concerning the true process.

The same problem has received considerable attention in the area of machine learning. We discuss how the observed instabilities in portfolio optimization (elaborated in Section \ref{instab}) can be understood and remedied by looking at portfolio theory from the point of view of machine learning.

Portfolio optimization is a special case of regression, and therefore can be understood as a machine learning problem (see Section \ref{reasons}). In machine learning, as well as in portfolio optimization, one wishes to minimize the {\em actual risk}, which is the risk (or error) evaluated by taking the ensemble average. This quantity, however, can not be computed from the data, only the {\em empirical risk} can. The difference between the two is not necessarily small in the thermodynamic limit, so that a small empirical risk does not automatically guarantee small actual risk \cite{VapnikCh71}. 

Statistical learning theory \cite{VapnikCh71, Vapnik95, Vapnik98}  finds upper bounds on the generalization error that hold with a certain accuracy. These error bounds quantify the expected generalization performance of a model, and they decrease with decreasing {\em capacity} of the function class that is being fitted to the data. Lowering the capacity therefore lowers the error bound and thereby improves generalization. The resulting procedure is often referred to as regularization and essentially prevents over-fitting (see Section \ref{RPO}).

In the thermodynamic limit, portfolio optimization needs to be regularized. We show in Section \ref{RPO-ES} how the above mentioned concepts, which find their practical application in support vector machines \cite{Boser92, CortesVapnik95}, can be used for portfolio optimization. Support vector machines constitute an extremely powerful class of learning algorithms which have met with considerable success. We show that regularized portfolio optimization, using the expected shortfall as a risk measure, is almost identical to support vector regression, apart from the budget constraint. We provide the modified optimization problem which can be solved by linear programming. 

In Section \ref{diversification}, we discuss the financial meaning of the regularizer: minimizing the L2 norm of the weight vector corresponds to a diversification pressure. We also discuss alternative constraints that could serve as regularizers in the context of portfolio optimization. 

Taking this machine learning angle allows one to organize a variety of ideas in the existing literature on portfolio optimization filtering methods into one systematic and well developed framework. There are basically two choices to be made: (i) which risk measure to use, and (ii) which regularizer. These choices result in different methods, because different optimization problems are being solved. 

While we focus here on the popular expected shortfall risk measure (in Section \ref{RPO-ES}), the variance has a long history as an important risk measure in finance. Several existing filtering methods that use the variance risk measure essentially implement regularization, without necessarily stating so explicitly. The only work we found in this context \cite{safePO} that mentiones regularization in the context of portfolio optimization has not been noticed by the ensuing, closely related, literature. It is easy to show that when the L2 norm is used as a regularizer, then the resulting method is closely related to Bayesian ridge regression, which uses a Gaussian prior on the weights (with the difference of the additional budget constraint). The work on covariance shrinkage, such as \cite{Jagannathan2003,LedoitWolf2003,LedoitWolf2004,LedoitWolfHoney}, falls into the same category. Other priors can be used \cite{DeMiguel2009}, which can be expected to lead to different results (for an insightful comparison see e.g. \cite{Tibshirani96}). Using the L1 norm has been popularized in statistics as the ``LASSO" (least absolute shrinkage and selection operator) \cite{Tibshirani96}, and methods that use any Lp norm are also known as the ``bridge" \cite{Frank93}. 

\section{Preliminaries -- Instability of classical portfolio optimization.}
\label{instab}
Portfolio optimization in large institutions operates in what we called the thermodynamic limit, where both the number of assets and the number of data points are large, with their ratio a certain, typically not very small, number. The estimation problem for the mean is so serious \cite{chopraziemba1993,merton1980} as to make the trade-off between risk and return largely illusory. Therefore, following a number of authors \cite{Jagannathan2003,LedoitWolf2003,okhrin2006,kempf_memmel2006,Frahm2008}, we focus on the minimum variance portfolio and drop the usual constraint on the expected return. This is also in line with previous work (see \cite{kondor2007} and references therein), and makes the treatment simpler without compromising the main conclusions. An extension of the results to the more general case is straightforward.

Nevertheless, even if we forget about the expected return constraint, the problem still remains that covariances have to be estimated from finite samples. It is an elementary fact from linear algebra that the rank of the empirical $N\times N$ covariance matrix is the smaller of $N$ and $T$. Therefore, if $T < N$, the covariance matrix is singular and the portfolio selection task becomes meaningless. The point $T = N$ thus separates two regions: for $T > N$ the portfolio problem has a solution, whereas for $T < N$, it does not.

Even if $T$ is larger than $N$, but not {\em much} larger, the solution to the minimum variance problem is unstable under sample fluctuations, which means that it is not possible to find the optimal portfolio in this way. This instability of the estimated covariances, and hence of the optimal solutions, has been generally known in the community, however, the full depth of the problem has only been recognized recently, when it was pointed out that the average estimation error diverges at the critical point $N = T$ \cite{pafka2002,pafka2003,pafka2004}.

In order to characterize the estimation error, Kondor and co-workers used the ratio $q_0^2$ between (i) the risk, evaluated at the optimal solution obtained by portfolio optimization using finite data and (ii) the true minimal risk. This quantity is a measure of generalization performance, with perfect performance when $q_0^2 = 1$, and increasingly bad performance as $q_0^2$ increases. As found numerically in \cite{pafka2003} and demonstrated analytically by random matrix theory techniques in \cite{burda2003}, the quantity $q_0$ is proportional to $(1 - N/T)^{-1/2}$ and diverges when $T$ goes to $N$ from above.

The identification of the point $N = T$ as a phase transition \cite{kondor2007,ciliberti_1} allowed for the establishment of a link between portfolio optimization and the theory of phase transitions, which helped to organize a number of seemingly disparate phenomena into a single coherent picture with a rich conceptual content. For example, it has been shown that the divergence is not a special feature of the variance, but persists under all the other alternative risk measures that have been investigated so far: historical expected shortfall, maximal loss, mean absolute deviation, parametric VaR, expected shortfall, and semivariance  \cite{kondor2007, ciliberti_1,ciliberti_2,hasszan2008}. The critical value of the $N/T$ ratio, at which the divergence occurs, depends on the particular risk measure and on any parameter that the risk measure may depend on (such as the confidence level in expected shortfall). However, as a manifestation of universality, the power law governing the divergence of the estimation error is independent of the risk measure \cite{ kondor2007, ciliberti_1, ciliberti_2}, the covariance structure of the market \cite{pafka2004}, and the statistical nature of the underlying process \cite{hasszan2007}. Ultimately, this line of thought led to the discovery of the instability of coherent risk measures \cite{kondor2008}.

\section{Statistical reasons for the observed instability in portfolio optimization}
\label{reasons}
As mentioned above, for simplicity and clarity of the treatment we do not impose a constraint on the expected return, and only look for the global minimum risk portfolio. This task can be formalized as follows: Given a fixed budget, customarily taken to be unity, given $T$ past measurements of the returns of $N$ assets: $x_i^k$, $i=1, \dots,N$, $k = 1, \dots, T$, and given the risk functional $F(\wx)$, find a weighted sum (the portfolio), $\wx$,\footnote{Notation: bold face symbols are understood to denote vectors.} such that it minimizes the {\em actual} risk
\begin{equation}
R (\w) = \langle F(\wx) \rangle_{p(\x)},
\end{equation}
under the constraint that $\sum_i w_i = 1$. The central problem is that one does not know the distribution $p(\x)$, which is assumed to underly the generation of the data. In practice, one then minimizes the {\em empirical} risk, replacing ensemble averages by sample averages:
\begin{equation}
R_{\rm emp} (\w) = {1 \over T} \sum_{k=1}^T F(\wxk)
\label{emprisk}
\end{equation}
Now, let us interpret the weight vector as a linear model. The model class given by the linear functions has a {\em capacity} $h$, which is a concept that has been introduced by Vapnik and Chervonenkis in order to measure how powerful a learning machine is \cite{VapnikCh71, Vapnik95, Vapnik98}.  (In the statistical learning literature, a learning machine is thought of as having a function class at its disposal, together with an induction principle and an algorithmic procedure for the implementation thereof \cite{Bernhard_thesis}). The capacity measures how powerful a function class is, and thereby also how easy it is to learn a model of that class. The rough idea is this: a learning machine has larger capacity if it can potentially fit more different types of data sets. Higher capacity comes, however, at the cost of potentially over-fitting the data. Capacity can be measured, for example, by the Vapnik-Chervonenkis (VC-) dimension \cite{VapnikCh71}, which is a combinatoric measure that counts how many data points can be separated in all possible ways by any function of a given class. 

To make the idea tangible for linear models, focus on two dimensions ($N=2$). For each number of points, $n$, one can choose the geometrical arrangement of the points in the plane freely. Once it is chosen, points are labeled by one of two labels, say ``red" and ``blue". Can a line separate the red points from the blue points for {\em any} of the $2^n$ different ways in which the points could be colored? The VC-dimension is the largest number of points for which this can be done. Two points can trivially be separated by a line. Three points that are not arranged collinear can still be separate for any of the 8 possible labelings. However, for four points this is no longer the case, since there is no geometrical arrangement for which one could not find a labeling that can not be separated by a line. The VC-dimension is 3, and in general, for linear models in $N$ dimensions, it is $N+1$ \cite{Bernhard_thesis, Bernhard_Book1}.

In the regime in which the number of data points are much larger than the capacity of the learning machine, $h/T << 1$, a small empirical risk guarantees small actual risk \cite{VapnikCh71}. For linear functions through the origin that are otherwise unconstrained, the VC-dimension grows with $N$. In the thermodynamic regime, where $N/T$ is not very small, minimizing the empirical risk does not necessarily guarantee a small actual risk \cite{VapnikCh71}. Therefore it is not guaranteed to produce a solution that generalizes well to other data drawn from the same underlying distribution. 

In solving the optimizing problem that minimizes the {\em empirical} risk, Eq. (\ref{emprisk}) in the regime in which $N/T$ is not very small, portfolio optimization {\it over-fits} the observed data. It thereby finds a solution that essentially pays attention to the seeming correlations in the data which come from estimation noise due to finite sample effects, rather than from real structure. The solution is thus different for different realizations of the data, and does not necessarily come close to the actual optimal portfolio. 

\section{Overcoming the instability}
\label{RPO}
The generalization error can be bounded from above (with a certain probability) by the empirical error plus a confidence term that is monotonically increasing with some measure of the capacity, and depends on the probability with which the bound holds \cite{Vapnik79}. Several different bounds have been established, connected with different measures of capacity, see e.g. \cite{Bernhard_Book1}. 

Poor generalization and over-fitting can be improved upon by decreasing the capacity of the model \cite{Vapnik95, Vapnik98}, which helps to lower the generalization error. Support vector machines are a powerful class of algorithms that implement this idea.

We suggest that if one wants to find a solution to the portfolio optimization problem in the thermodynamic regime, then one should not minimize the empirical risk alone, but also constrain the capacity of the portfolio optimizer (the linear model). 

How can portfolio optimization be regularized? Portfolio optimization is essentially a regression problem, and therefore we can apply statistical learning theory, in particular the work on support vector regression. 

Note first that the capacity of a linear model class for which the length of the weight vector is restricted to $\|w\|^2 \leq A$ has an upper bound which is smaller than the capacity of unconstrained linear models \cite{Vapnik95, Vapnik98}. The capacity is minimized when the length of the weight vector is minimized \cite{Vapnik95, Vapnik98}. Vapnik's concept of {\em structural risk minimization} \cite{Vapnik79} results in the support vector algorithm \cite{Boser92, CortesVapnik95} which finds the model with the smallest capacity that is consistent with the data, that is the model with smallest $\|w\|^2$. This leads to a convex constrained optimization problem \cite{Boser92, CortesVapnik95} which can be solved using linear programming. 

\section{Regularized portfolio optimization with the expected shortfall risk measure.}
\label{RPO-ES}
While the original Markowitz' formulation \cite{Markowitz52} measures
risk by the variance, many other risk measures have been proposed
since. Today, the most widely used risk measure, both in practice and in
regulation, is Value at Risk (VaR) \cite{jorion,riskmetrics}. VaR has,
however, been criticized for its lack of convexity, see e.g.
\cite{Artzner99, embrechts,acerbi1}, and an axiomatic approach,
leading to the introduction of the class of coherent risk measures,
was put forward \cite{Artzner99}. Expected shortfall, essentially a
conditional average measuring the average loss above a high threshold,
has been demonstrated to belong to this class
\cite{acerbi2,acerbi3,acerbi4}.

Expected shortfall has been steadily gaining popularity in recent years. The regularization we propose here is intended to cure its weak point, the sensitivity to sample fluctuations, at least for reasonable values of the ratio $N/T$. 

Choose the risk functional $F(z) = z \theta(z-\alpha_\beta)$, where
$\alpha_\beta$ is a threshold, such that a given fraction $\beta$ of the
(empirical) loss-distribution over $z$ lies above $\alpha_\beta$. One
now wishes to minimize the average over the remaining tail
distribution, containing the fraction $\nu :=1 - \beta$, and
defines the expected shortfall as
\begin{equation}
ES = \min_{\epsilon}\left[\epsilon + \frac{1}{\nu T} \sum_{k=1}^T {1
\over 2} \left(-\epsilon -\wxk + |-\epsilon -\wxk|\right)\right].
\label{ExpS}
\end{equation}
The term in the sum implements the $\theta$-function, while $\nu$ in
the denominator ensures normalization of the tail distribution. It has
been pointed out \cite{Rockafellar} that this optimization problem
maps onto solving the linear program:
\begin{eqnarray}
&&\min_{\w, {\bf \xi}, \epsilon}  \left[ {1 \over T} \sum_{k=1}^{T}
\xi_k + \nu \epsilon \right] \label{CVaR} \\
&{\rm s.t.}\;\;\; & \wxk + \epsilon + \xi_k \geq 0; \;\;\; \xi_k; \geq
0 \label{ES-constr} \\
&&  \sum_i w_i  = 1.
\end{eqnarray}
We propose to implement regularization by including the minimization of $\|\w \|^2 $. This can be done using a Lagrange multiplier, $C$, to control the trade-off -- as we relax the constraint on the length of the weight vector, we can, of course, make the empirical error go to zero and retrieve the solution to the minimal expected shortfall problem. The new optimization problem reads:

\begin{eqnarray}
&&\min_{\w, {\bf \xi}, \epsilon}  \left[ {1 \over 2}  \|\w \|^2 +  C
\left({1 \over T} \sum_{k=1}^{T} \xi_k + \nu \epsilon \right) \right]
\label{newPO} \\
&{\rm s.t.}\;\;\; & - \wxk \leq \epsilon + \xi_k; \label{con1}\\
&&  \xi_k \geq 0; \;\;\; \epsilon \geq 0;\\
&& \sum_i w_i = 1. \label{b}
\end{eqnarray}
The problem is mathematically almost identical to a support vector regression (SVR) algorithm called $\nu$-SVR. There are two differences: (i) the budget constraint is added, and (ii) the loss function is asymmetric. Expected shortfall is an asymmetric version of the $\epsilon$-intensive loss, used in support vector regression, defined as the maximum of $\{0;| f(\x) - y| - \epsilon \}$, where $f(\x)$ is the interpolant, and $y$ the measured value (response). In that sense $\epsilon$ measures an allowable error below which deviations are discarded.\footnote{The mathematical similarity between minimum expected shortfall {\em without} regularization and the E$\nu$-SVM algorithm \cite{EnuSVM} was pointed out, but incorrectly, in \cite{Takeda2008}. There is an important difference between the two optimization problems. In E$\nu$-SVM, the length of the weight vector, $\| \w \|$, is constrained, which implements capacity control. In the pure expected shortfall minimization, Eq. (\ref{CVaR}), this is not done. Instead, the total
budget $\sum_i w_i$ is fixed. This difference is not correctly identified in the proof of the central theorem (Theorem 1) in \cite{Takeda2008}.}

The use of asymmetric risk measures in finance is motivated by the consideration that investors are not afraid of upside fluctuations. However, to make the relationship to support vector regression as clear as possible, we will first solve the more general symmetrized problem, before restricting our treatment to the completely asymmetric case, corresponding to expected shortfall. In addition, one may argue that focusing exclusively on large negative fluctuations might not be advisable even from a financial point of view, especially when one does not have sufficiently large samples. In a relatively small sample it may happen that a particular item, or a certain combination of items, dominates the rest, i.e. produces a larger return than any other item in the portfolio at each time point, even though no such dominance exists on longer time scales. The probability of such an apparent arbitrage increases with the ratio $N/T$, and when it occurs it may encourage an investor acting on a lopsided risk measure to take up very large long positions in the dominating item(s), which may turn out to be detrimental on the long run. This is the essence of the argument that has led to the discovery of the instability of coherent and downside risk measures \cite{hasszan2008,kondor2008}.

According to the above, let us consider the general case where positive deviations are also penalized. The objective function, Eq. (\ref{newPO}), then becomes
\begin{equation}
\min_{\w, {\bf \xi}, \epsilon}  \left[ {1 \over 2}  \|\w \|^2 +  C
\left({1 \over T} \sum_{k=1}^{T} \left( \xi_k + \xi_k^*\right) + \nu
 \epsilon \right) \right] \label{gen-RES},
\end{equation}
and additional constraints have to be added to Eqs. (\ref{con1}) to (\ref{b}):
\begin{eqnarray}
\wxk \leq \epsilon + \xi_k^*; \;\;\; \xi_k^* \geq 0.\label{con1sym}
\end{eqnarray}
This problem corresponds to $\nu$-SVR, a well understood regression method \cite{Nu-SVM}, with the only difference that the budget constraint, Eq. (\ref{b}) is added here. In the finance context the associated loss might be called {\em symmetric tail average} (STA). Solving the regularized expected shortfall minimization problem, Eqs. (\ref{newPO})--(\ref{b}) is a special case of solving the regularized STA minimization problem, Eq. (\ref{gen-RES}) with the constraints Eqs. (\ref{con1})--(\ref{b}) and (\ref{con1sym}). Therefore, we solve the more general problem first (Section \ref{RSTA}), before providing, in Section \ref{RES-par}, the solution to the regularized expected shortfall, Eqs. (\ref{newPO})--(\ref{b}).

\subsection{Regularized Symmetric Tail Average Minimization}
\label{RSTA}
The solution to the regularized symmetric tail average problem, Eq. (\ref{gen-RES}) with the constraints Eqs. (\ref{con1})--(\ref{b}) and (\ref{con1sym}), is found in analogy to support vector regression, following \cite{Nu-SVM}, by writing down the Lagrangean,
using Lagrange multipliers, $\{ {\bf \alpha}, {\bf \alpha^*}, \gamma, \lambda, {\bf \eta}, {\bf \eta^*} \}$, for the constraints. The
solution is then a saddle point, i.e. minimum over primal and maximum over dual variables. The Lagrangean is different from the one that arises in $\nu$-SVR in
that it is modified by the budget constraint: 
\begin{eqnarray}
\!\!\!\!\! \!\!\!\!\! \!\!\!\!\! \!\!\!\!\! \!\!\!\!\! \!\!\!\!\! \!\!\!\!\!  L[\w, {\bf \xi}, {\bf \xi^*}, \epsilon, {\bf \alpha}, {\bf \alpha^*}, \gamma, \lambda, {\bf \eta}, {\bf \eta^*}]
&=&{1 \over 2} \|\w \|^2 + {C \over T} \sum_{k=1}^{T} (\xi_k +
\xi_k^*) + C \nu \epsilon  - \lambda \epsilon + \gamma \left( \sum_i
w_i -1 \right)\nonumber \\
&& + \sum_{k=1}^{T} \alpha_k^* (\wxk - \epsilon - \xi_k^*) -
\sum_{k=1}^{T} \alpha_k (\wxk + \epsilon + \xi_k)
 \nonumber \\
&& - \sum_{k=1}^{T} (\eta_k  \xi_k + \eta_k^*  \xi_k^*) \label{L-sym} \\
&=& F[\w] + \epsilon  \left( C \nu - \lambda -  \sum_{k=1}^{T}
(\alpha_k + \alpha_k^*) \right) -\gamma \label{Lagr-w-constr} \\ &&+
\sum_{k=1}^{T} \left[ \xi_k \left( {C \over T} -\alpha_k -
\eta_k\right) + \xi_k^* \left( {C \over T} - \alpha_k^* - \eta_k^*
\right) \right] \nonumber
\label{L1}
\end{eqnarray}
with
\begin{eqnarray}
F[\w] &=& \w \cdot \left({1 \over 2}  \w - \left(\sum_{k=1}^{T}
(\alpha_k - \alpha_k^*) \xk - \gamma {\bf 1}\right) \right),
\end{eqnarray}
where ${\bf 1}$ denotes the unit vector of length $N$.
Setting the derivative of the Lagrangian w.r.t. $\w$ to zero gives:
\begin{eqnarray}
\w_{\rm opt} = \sum_{k=1}^{T} (\alpha_k - \alpha_k^*) \xk - \gamma
{\bf 1} \label{w-sym}
\end{eqnarray}
This solution for the optimal portfolio is sparse in the sense that,
due to the Karush-Kuhn-Tucker conditions (see e.g. \cite{Bertsekas95}), only those points contribute to the
optimal portfolio weights, for which
the inequality constraints in (\ref{con1}), and the corresponding constraints in Eq. (\ref{con1sym}), are met exactly. The solution of $\w_{\rm opt} $ contains
only those points, and effectively ignores the rest. This
sparsity contributes to the stability of the solution.
Regularized portfolio optimization (RPO) operates, in contrast to
general regression, with a fixed budget. As a consequence, the
Lagrange multiplier $\gamma$ now appears in the optimal solution, Eq.
(\ref{w-sym}). Compared to the optimal solution in support vector (SV) regression, $\w_{\rm SV}$, the solution
vector under the budget constraint, $\w_{\rm RPO}$, is shifted by
$\gamma$:
\begin{equation}
\w_{\rm RPO} = \w_{\rm SV} - \gamma {\bf 1}.
\end{equation}
Let us now consider the dual problem. The dual is, in general, a
function of the dual variables, which are here $\{ {\bf \alpha}, {\bf
\alpha^*}, \gamma, \lambda, {\bf \eta}, {\bf \eta^*} \}$, although we
will see in the following that some of these variables drop out. The
dual is defined as \mbox{$D := \min_{\w, {\bf \xi}, {\bf \xi^*},
\epsilon} L[\w, {\bf \xi}, {\bf \xi^*}, \epsilon, {\bf \alpha}, {\bf
\alpha^*}, \gamma, \lambda, {\bf \eta}, {\bf \eta^*}]$}, and the dual
problem is then to maximize $D$ over the dual variables. We can
replace the minimization over $\w$ by evaluating the Lagrangian at
$\w_{\rm opt}$. For that we have to evaluate
\begin{eqnarray}
F[\w_{\rm opt}] &=& - {1 \over 2}  \| \w_{\rm opt} \|^2 \\
&=& \left[- {1 \over 2}
\left(\sum_{k=1}^{T} (\alpha_k - \alpha_k^*) \xk - \gamma {\bf
1}\right)^2 \right].
\end{eqnarray}
For the other terms in the Lagrangian, we have to consider different cases:
\begin{enumerate}
\item If $\left( C \nu - \lambda -  \sum_{k=1}^{T} (\alpha_k +
\alpha_k^*) \right) < 0$, then $L$ can be minimized by letting
$\epsilon \rightarrow \infty$, which means that $D = -\infty$.
\item If $\left( C \nu - \lambda -  \sum_{k=1}^{T} (\alpha_k +
\alpha_k^*) \right) \geq 0$: The term $\epsilon \left( C \nu - \lambda
-  \sum_{k=1}^{T} (\alpha_k + \alpha_k^*) \right)$ vanishes. Reason:
if equality holds, this is trivially true, and if the inequality holds
strictly then $L$ can be minimized by setting $\epsilon =0$.
\end{enumerate}
Similarly, for the other constraints (the notation $(*)$ means that
this is true for variables with and without the asterisk):
\begin{enumerate}
\item If $\left( {C \over T} -\alpha_k^{(*)} - \eta_k^{(*)} \right) <
0$, then $L$ can be minimized by letting $\xi_k^{(*)} \rightarrow
\infty$, which means that $D = -\infty$.
\item If $\left( {C \over T} -\alpha_k^{(*)} - \eta_k^{(*)} \right)
\geq 0$, then $\xi_k \left( {C \over T} -\alpha_k^{(*)} - \eta_k^{(*)}
\right) = 0$. Reason: If the inequality holds strictly then $L$ can be
minimized by $\xi_k^{(*)} = 0$. If equality holds then it is trivially
true.
\end{enumerate}
By a similar argument, the term $\gamma$ in Eq. (\ref{Lagr-w-constr})
disappears in the Dual. Altogether we have that either $D = - \infty$,
or
\begin{eqnarray}
&& D({\bf \alpha}, {\bf \alpha^*}, \gamma) = \min_{{\bf \xi}, {\bf
\xi^*}, \epsilon} F[\w_{\rm opt}({\bf \alpha}, {\bf \alpha^*},
\gamma)] =  - {1 \over 2}  \| \w_{\rm opt} \|^2  \\
{\rm and} &\;\;\;& \sum_{k=1}^{T} (\alpha_k^* + \alpha_k) \leq C \nu -
\lambda \\
{\rm and} &\;\;\;& \alpha_k^{(*)} + \eta_k^{(*)} \leq {C \over T}.
\end{eqnarray}
Note that the variables $\xi_k^{(*)}, \eta_k^{(*)}, \epsilon, \lambda$
do not appear in $F[\w_{\rm opt}({\bf \alpha}, {\bf \alpha^*},
\gamma)]$. The dual problem is therefore given by
\begin{eqnarray}
\max_{{\bf \alpha}, {\bf \alpha^*}, \gamma} && \left[- {1 \over 2}
\left(\sum_{k=1}^{T} (\alpha_k - \alpha_k^*) \xk - \gamma {\bf
1}\right)^2 \right]. \\
{\rm s.t.}&& \{ \alpha_k, \alpha_k^* \} \in \left[0,{C \over T}\right] \\
&& \sum_{k=1}^{T} (\alpha_k^* + \alpha_k) \leq C\nu.
\end{eqnarray}
We can analytically maximize over $\gamma$ and obtain for the optimal value
\begin{equation}
\gamma = {1 \over N} \left(\sum_{k=1}^{T} (\alpha_k - \alpha_k^*)
\sum_{i=1}^N \xki - 1 \right)  \label{gamma}
\end{equation}
The optimal projection (= optimal portfolio) is given by
\begin{equation}
\!\!\!\!\!\!\!\! \!\!\!\! \!\!\!\! \!\!\!\! \!\!\!\!  \w_{\rm opt} \cdot \x = \sum_{k=1}^{T}  (\alpha_k - \alpha_k^*)  \xk \cdot \x - {1 \over N}
\left(\sum_{k=1}^{T}  (\alpha_k - \alpha_k^*)  \sum_{i=1}^N  \xki - 1
\right) {\bf1} \cdot \x .
\label{RPO-sol-w}
\end{equation}
For $N \rightarrow \infty$ the second term vanishes and the solution is the same as the the solution in support vector regression. Note that the kernel-trick (see e.g. \cite{Bernhard_Book1}), which is
used in support vector machines to find nonlinear models hinges on the
fact that only dot products of input vectors appear in the support vector
expansion of the solution. As a consequence
of the budget constraint, one can no longer use the kernel-trick
(compare Eq. (\ref{RPO-sol-w})). As long as we disregard derivatives, this is
not a problem for portfolio optimization. Keep in mind, however, that the budget
constraint introduces this otherwise undesirable property.

Support vector algorithms typically solve the dual form of the problem
(for a recent survey see \cite{leon2006}), which is in our case given
by
\begin{eqnarray}
\!\!\!\!\!\!\!\! \max_{{\bf \alpha}, {\bf \alpha^*}, \gamma} && - {1 \over 2} \left[
\sum_{k=1}^{T}  \sum_{l=1}^{T} (\alpha_k - \alpha_k^*)(\alpha_l -
\alpha_l^*) \left( \xk \xl - {1\over N} \sum_{i=1}^{N} x^{(k)}_i
\sum_{i=1}^{N} x^{(l)}_i \right) 
\right] \label{dual-sym-rpo} \\
\!\!\!\!\!\! {\rm s.t.}&& \{ \alpha_k, \alpha_k^* \} \in \left[0,{C \over T}\right]; \nonumber \\
&& \sum_{k=1}^{T} (\alpha_k^* + \alpha_k) \leq C\nu. \nonumber
\end{eqnarray}
For $N \rightarrow \infty$ the problem becomes {\it identical} to
$\nu$-SVR, which can be solved by linear programming, for
which software packages are available \cite{LOQO}. For finite $N$, it can still be solved with existing
methods, because it is quadratic in the $\alpha_k$'s. Solvers such as
the ones discussed in \cite{bordes-ertekin-weston-bottou-2005} and
\cite{leon2006} can be used, but have to be adapted to this specific problem.

The regularized symmetric tail average minimization problem (Eq. (\ref{gen-RES}) with the constraints Eqs. (\ref{con1})--(\ref{b}) and (\ref{con1sym})) is, as we have shown here, directly
related to support vector regression which uses the $\epsilon$-insensitive loss function. The $\epsilon$-insensitive
loss is stable to local changes for data points that fall outside the range specified by $\epsilon$. This
point is elaborated in Section 3 in \cite{Nu-SVM}, and relates this
method to robust estimation of the mean. It can also be extended to robust estimation of quantiles \cite{Nu-SVM} by scaling
of the slack variables $\xi_k$ by $\mu$ and $\xi_k^*$ by $1-\mu$, respectively.

This scaling translates directly to the portfolio optimization
problem, which is an extreme case: downside risk measures 
penalize only loss, not gain. The asymmetry in the loss function
corresponds to $\mu =1$.

\subsection{Regularized expected shortfall.}
\label{RES-par}
By this final change we arrive at the regularized portfolio
optimization problem, Eqs. (\ref{newPO})--(\ref{b}), which we
originally set out to solve. This is now easily solved in analogy to
the previous paragraphs: the slack variables $\xi_k^*$ disappear,
together with the respective Lagrange multipliers which enforce
constraints, including $\alpha_k^*$. The optimal solution is now
\begin{eqnarray}
&\w_{\rm opt} = \sum_{k=1}^{T} \alpha_k  \xk - \gamma {\bf1},
\end{eqnarray}
with
\begin{eqnarray}
\gamma &=& {1 \over N} \left(\sum_{k=1}^{T} \alpha_k \sum_{i=1}^{N}
\xki - 1 \right).
\end{eqnarray}
The dual problem is given by
\begin{eqnarray}
\max_{\alpha_k}&& - {1 \over 2} \left[ \sum_{k=1}^{T}  \sum_{l=1}^{T}
\alpha_k \alpha_l \left( \xk \xl - {1\over N} \sum_{i=1}^{N} x^{(k)}_i
\sum_{i=1}^{N} x^{(l)}_i \right) 
\right] \nonumber \\
{\rm s.t.}&& \alpha_k \in \left[0,{C \over T}\right]; \;\;
\sum_{k=1}^{T} \alpha_k \leq C\nu.
\end{eqnarray}
which, like its symmetric counterpart, Eq. (\ref{dual-sym-rpo}), can
be solved by adjusting existing algorithms.

The formalism provides a free parameter, $C$, to set the balance between the original risk function and the regularizer. Its choice may depend on a number of factors, such as the investors time horizon, the nature of the underlying data, and, crucially, on the ratio $N/T$. Intuitively, there must be a maximum allowable value $C_{\rm max}(N/T)$ for $C$, such that when one puts more emphasis on the data, $C > C_{\rm max}(N/T)$, then over fitting will occur with high probability. It would be desirable to know an analytic expression for (a bound on) $C_{\rm max}(N/T)$. In practice, cross-validation methods are often employed in machine learning to set the value of $C$. Those methods are not free of problems (see, for example, the treatment in \cite{bengio}), and the optimal choice of this parameter remains an open problem. 

\section{Regularization corresponds to portfolio diversification.}
\label{diversification}
Above, we have controlled the capacity of the linear model by minimizing the L2 norm of the portfolio weight vector. In the finance context, minimizing 
\begin{equation}
\| \w \|^2 = \sum_i w_i^2 \simeq {1 \over N_{\rm eff}}
\end{equation}
corresponds roughly to maximizing the effective number of assets, $N_{\rm eff}$, i.e. to exerting a {\it pressure} towards portfolio diversification \cite{Bouchaudpotters}. We conclude that diversification of the portfolio is crucial, because it serves to counteract the observed instability by acting as a regularizer. 

Other constraints that penalizes the length of the weight vector could alternatively be considered as a regularizer, in particular any Lp norm. The budget constraint {\em alone}, however, does not suffice as a regularizer, since it does not constrain the length of the weight vector. 
Adding a ban on short selling, $w_i \geq 0$, to the budget constraint, $\sum_i w_i = 1$, limits the allowable solutions to a finite volume in the space of weights and is equivalent to requiring that $\sum_i | w_i | \leq 1$.\footnote{This point has been made independently by \cite{DeMiguel2009}.} It thereby imposes a limit on the L1 norm, that is on the sum of the absolute amplitudes of long and short positions. 

One may argue that it may be a good idea to use the L1 norm instead of the L2 norm, because that may make the solution sparser. However, the L1 norm has a tendency to make some of the weights vanish. Indeed, it has been shown that in the orthonormal design case (using the variance as the risk measure) an L1 regularizer will set some of the weights to zero, while an L2 regularizer will scale all the weights \cite{Tibshirani96}. The spontaneous reduction of portfolio size has also been demonstrated in numerical simulations \cite{Kondor4}: as one goes deeper and deeper into the regime where $T$ is significantly smaller than $N$, under a ban on short selling, more and more of the weights will become zero. The same ``freezing out" of the weights has been observed in portfolio optimization \cite{scherermartin} as an empirical fact.  

It is important to stress that the vanishing of some of the weights does not reflect any structural property of the objective function, it is  just a random effect: as clearly demonstrated by simulations \cite{Kondor4}, for a different sample a different set of weights vanishes. The angle of the weight vector fluctuates  wildly from sample to sample. (The behavior of the solutions is similar for other limit systems as well.) This means that the solutions will be determined by the limit system and the random sample, rather than by the structure of the market. So the underlying instability is merely ``masked", in that the solutions do not run away to infinity, but they are still unstable under sample fluctuations when $T$ is too small. As it is certainly not in the interest of the investor to obtain a portfolio solution which sets weights to zero on the basis of unreliable information from small samples, the above observations speak strongly in favor of using the L2 norm over the L1 norm. 

\section{Conclusion}
We have made the observation that the optimization of large portfolios minimizes the empirical risk in a regime where the data set size is similar to the size of the portfolio. In that regime, a small empirical risk does not necessarily guarantee a small actual risk \cite{VapnikCh71}. In this sense naive portfolio optimization over-fits the data. Regularization can overcome this problem by reducing the capacity of the considered model class. 

Regularized portfolio optimization has choices to make, not only about the risk function, but also about the regularizer. Here, we have focussed on the increasingly popular expected shortfall risk measure. Using the L2 norm as a regularizer leads to a convex optimization problem which can be solved with linear programming. We have shown that regularized portfolio optimization is then a variant of support vector regression. The differences are an asymmetry, due to the tolerance to large positive deviations, and the budget constraint, which is not present in regression. 

Our treatment provides a novel insight into why diversification is so important. The L2 regularizer implements a pressure towards portfolio diversification. Therefore, from a statistical point of view, diversification is important as it is one way to control the capacity of the portfolio optimizer and thereby to find a solution which is more stable, and hence meaningful. 

In summary, the method we have outlined in this paper allows for the unified treatment of optimization and diversification in one principled formalism. It shows how known methods from modern statistics can be used to improve the practice of portfolio optimization.

\section{Acknowledgements}
We thank Leon Bottou for helpful discussions and comments on the manuscript. This work has been supported by the ``Cooperative Center for Communication Networks Data Analysis", a NAP project sponsored by the National Office of Research and Technology under grant No. KCKHA005. SS thanks the Collegium Budapest for hosting her during this collaboration, and the community at the Collegium for providing a creative and inspiring atmosphere.

\bibliography{RPOsubmit}

\end{document}